\documentclass{book}

\usepackage{amsmath,latexsym,amssymb,kcp,named}

\newcommand{\ket}[1]{\ensuremath{|#1\rangle}}
\newcommand{\bra}[1]{\ensuremath{\langle #1|}}
\newcommand{\scprod}[2]{\ensuremath{\langle #1 | #2 \rangle }}
\newcommand{\colvec}[1]{
   \left ( \begin{array}{c}
              #1_1 \\
              \vdots \\
              #1_n
              \end{array}
    \right ) 
}      
\newcommand{\lenvec}[1]{\ensuremath{\|#1\|}}

\newcommand\remove[1]{}

\begin{document}

\paper[Quantum Logic of Semantic Space]%
{Quantum Logic of Semantic Space: an exploratory investigation of context effects
  in practical reasoning
}%
{
  P. D. Bruza and R. J. Cole
}

%\title{Quantum Logic of Semantic Space}

\index{Bruza, P. D.}
\index{Cole, R. J.}

\section{Introduction}

The field of non-monotonic reasoning (NMR) has successfully provided an
impressive symbolic account of human practical reasoning over the last
two and half decades. There remains, however, a disappointment -
the dearth of large-scale operational NMR systems on the
ground. During Lora Morgenstern's keynote address at the International
Joint Conference on Artificial Intelligence (IJCAI-97) with the title
``Inheritance Comes of Age: Applying Non-monotonic Techniques to
Problems in Industry'' she warned researchers that NMR needs to go
beyond the examination of toy examples and to tackle serious, large scale
problems, or run the risk of NMR becoming a backwater at artificial intelligence
conferences. That is getting on ten years ago. Since then NMR has
largely crystallized and is well understood from a stratum of
theoretical perspectives.
Morgenstern's warning still lingers, in our opinion. Theoretical insight 
without corresponding reasoning systems on the ground belies NMR's promise 
of embodying human practical reasoning.

We feel that the symbolic characterization of practical reasoning is
only part of the picture. G\"{a}rdenfors
(~\cite{Book:00:Gardenfors:ConSpace}, p127) argues that one must go
under the symbolic level of cognition. In this vein, he states,
``\dots information about an object may be of two kinds: {\em
propositional} and {\em conceptual}. When the new information is
propositional, one learns new {\em facts} about the object, for
example, that $x$ is a penguin. When the new information is
conceptual, one {\em categorizes} the object in a new way, for
example, $x$ is {\em seen as} a penguin instead of as just a bird''.
G\"{a}rdenfors' mention of ``conceptual'' refers to the conceptual
level of a three level model of
cognition~\cite{Book:00:Gardenfors:ConSpace}.  How information is
represented varies greatly across the different levels. The
sub-conceptual level is the lowest level within which information is
carried by a connectionist representation. Within the uppermost level
information is represented symbolically. It is the intermediate,
\emph{conceptual level}, or \emph{conceptual space}\index{space,
conceptual}, which is of particular relevance to this account. Here
properties and concepts have a geometric representation in a
dimensional space.  For example, the property of ``redness'' is
represented as a convex region in a tri-dimensional space determined
by the dimensions hue, chromaticity and brightness.  The point left
dangling for the moment is that representation at the conceptual level
is rich in associations, both explicit and implicit.  We speculate
that the dynamics of associations are primordial stimuli for practical
inferences drawn at the symbolic level of cognition.  For example, it
seems that associations and analogies generated within conceptual
space play an important role in hypothesis generation. G\"{a}rdenfors
(\cite{Book:00:Gardenfors:ConSpace}, p48) alludes to this point when
he states, ``most of scientific theorizing takes place within the
conceptual level.'' His conjecture is aligned with Gabbay and Woods'
insights regarding the cognitive economic basis of
abduction~\cite{Book:05:Gabbay:Abduction}. Put crudely, it is cheaper
to ``guess'' than to pursue a deductive agenda in relation to a
problem at hand.  Gabbay and Woods' notion of cognitive economy rests
on compensation strategies employed by a practical agent to alleviate
the consequences of key cognitive resources such as {\em information},
{\em time}, and {\em computational capacity}.  Practical reasoning is
reasoning performed by practical agents, and is therefore subject to
cognitive economy. In this connection, we put forward the following
conjecture: It may well be that because such associations are formed
below the symbolic level of cognition, significant cognitive economy
results.  This is not only interesting from a cognitive point of view,
but also opens the door to providing a computationally tractable
practical reasoning systems, for example, operational abduction to
drive scientific discovery in biomedical
literature~\cite{Article:04:Bruza:Abd,Article:05:Bruza:Abduction}

The appeal of G\"{a}rdenfors' cognitive model is that it allows
inference to be considered not only at the symbolic level, but also at
the conceptual (geometric) level.  Inference at the symbolic level is
typically a linear, deductive process.  Within a conceptual space,
inference takes on a decidedly associational character because
associations are often based on similarity (e.g., semantic or
analogical similarity), and notions of similarity are naturally
expressed within a dimensional space.  For example, G\"{a}rdenfors'
states that a more natural interpretation of ``defaults'' is to view
them as ``relations between concepts''. This is a view which flows into
the account which follows: the strength of associations between concepts change dynamically under the influence of context.
This, in turn, influences the defaults haboured within the symbolic level of cognition.

It is important to note the paucity of representation at the symbolic
level and reflect how symbolic reasoning systems are hamstrung as a
result. In this connection, G\"ardenfors
(\cite{Book:00:Gardenfors:ConSpace}, p127) states, `` ..information
about categorization can be quite naturally transfered to
propositional information: categorizing $x$ as an emu, for example,
can be expressed by the proposition ``$x$ is an emu''. This
transformation into the propositional form, however, tends to suppress
the internal {\em structure} of concepts. Once one formalizes
categorizations of objects by {\em predicates} in a first-order
language, there is a strong tendency to view the predicates as
primitive, atomic notions and to forget that there are rich relations
among concepts that disappear when put into standard logical
formalism.''

The above contrast between the conceptual and symbolic levels raises
the question as to what are the implications for providing an account
of practical reasoning.  G\"{a}rdenfors states that concepts generate
``expectations that result in different forms of \emph{non-monotonic
reasoning}'', which are summarized as follows:

\subsubsection*{Change from a general category to a subordinate}

When shifting from a basic category, e.g., ``bird" to a subordinate
category, e.g., ``penguin", certain default associations are given up
(e.g., ``Tweety flies''), and new default properties may arise (e.g.,
``Tweety lives in Antarctica'').

\subsubsection*{Context effects} 

The context of a concept triggers different associations that ``lead
to non-monotonic inferences''. For example, {\em Reagan} has default
associations ``Reagan is a president'', ``Reagan is a republican'' etc.,
but {\em Reagan} seen in the context of {\em Iran} triggers associations of
``Reagan'' with ``arms scandal'', etc.

\subsubsection*{The effect of contrast classes}

Properties can be relative, for example, ``a tall Chihuahua is not a
tall dog'' (\cite{Book:00:Gardenfors:ConSpace}, p119).  In the first
contrast class ``tall'' is applied to Chihuahuas and the second
instance it is applied to dogs in general.  Contrast classes generate
conceptual subspaces, for example, skin colours form a subspace of the
space generated by colours in general.  Embedding into a subspace
produces non-monotonic effects.  For example, from the fact that $x$ is
a white wine and also an object, one cannot conclude that $x$ is a
white object (as it is yellow).

\subsubsection*{Concept combination}

Combining concepts results in non-monotonic effects.  For example,
\emph{metaphors} (\cite{Book:00:Gardenfors:ConSpace}, p130) Knowing that
something is a lion usually leads to inferences of the form that it is
alive, that it has fur, and so forth. In the combination, {\em stone
lion}, however, the only aspect of the object that is lion-like is its
shape. One cannot conclude that a stone lion has the other usual
properties of a lion, and thus we see the non-monotonicity of the
combined concept.

An example of the non-monotonic effects of concept combination not involving metaphor is the following: 
\emph{
A guppy
is not a typical pet, nor is guppy is a typical fish, but a guppy is a
typical pet fish}.

In short, concept combination leads to conceptual change. These
correspond to revisions of the concept and parallel belief revisions
modelled at the symbolic level, the latter having received thorough
examination in the artificial intelligence literature.

The preceding brief characterization of the dynamics of concepts and
associated non-monotonic effects is intended to leave the impression
that a lot of what is happening in relation with practical reasoning
is taking place within a conceptual (geometric) space.  In addition,
this impression may provide a foothold towards realizing genuine
operational systems. This would require at least three issues to be
addressed. The first is that a computational variant of the conceptual
level of cognition is required. Secondly, the non-monotonic effects
surrounding concepts would need to be formalized and
implemented. Thirdly, the connection between these effects and NMR at
the symbolic level needs to be specified. This account will attempt to
address the first two of these questions. Computational approximations
of conceptual space will be furnished by semantic space models which
are emerging from the fields of cognition and computational
linguistics.  Semantic space models not only provide a cognitively
motivated basis to underpin human practical reasoning, but from a
mathematical perspective, they are real-valued Hilbert spaces. This
introduces the tantalizing and highly speculative prospect of
formalizing aspects of human practical reasoning via quantum
mechanics\index{quantum mechanics}.  In this account will focus on a treatment
of how to formalize context effects as well as keeping an eye on
operational issues.

\section[Semantic Space]{
  Semantic space: computational approximations of conceptual space
}

To illustrate how the gap between cognitive knowledge representation
and actual computational representations, the Hyperspace Analogue to
Language (HAL)\index{hyperspace analogue to language} model is
employed~\cite{Article:96:Lund:HAL,Article:98:Burgess:HAL}.  HAL
produces representations of words in a high dimensional space that
seem to correlate with the equivalent human representations.  For
example, ``...simulations using HAL accounted for a variety of
semantic and associative word priming effects that can be found in the
literature...and shed light on the nature of the word relations found
in human word-association norm data''\cite{Article:96:Lund:HAL}.
Given an $n$-word vocabulary, a HAL space is an $n \times n$ matrix
constructed by moving a window of length $l$ over the corpus by one
word increment ignoring punctuation, sentence and paragraph
boundaries.  All words within the window are considered as
co-occurring with the last word in the window with a strength
inversely proportional to the distance between the words. Each row
$i$ in the matrix represents accumulated weighted associations of word
$i$ with respect to other words which preceded $i$ in a context
window.  Conversely, column $i$ represents accumulated weighted
associations with words that appeared after $i$ in a window.  For
example, consider the text ``President Reagan ignorant of the arms
scandal'', with $l=5$, the resulting HAL matrix $H$ would be:
\begin{table}[h!]
\begin{center}
\begin{tabular}{|c|c|c|c|c|c|c|c|} \hline
 & arms & ig & of & pres &  reag & scand & the \\ \hline
 arms & 0 & 3 & 4 & 1& 2& 0 & 5 \\ \hline
 ig & 0 & 0 & 0 & 4 & 5 & 0 & 0 \\ \hline
 of & 0 & 5 & 0 & 3 & 4 & 0 & 0 \\ \hline
 pres & 0 & 0 & 0 & 0 & 0 & 0 & 0 \\ \hline
 reag & 0 & 0 & 0 & 5 & 0 & 0 & 0 \\ \hline
 scand & 5 & 2 & 3 & 0 & 1 & 0 & 4 \\ \hline
the & 0 & 4& 5 & 2 & 3 & 0 & 0 \\ \hline  
\end{tabular}
\caption{A simple semantic space computed by HAL}
\end{center}
\label{table:reaganeg}
\end{table}
\begin{figure}
\begin{center} \begin{minipage}{10cm}{\small\tt
\begin{tabbing}
def calculate\_hal(documents, n) \\
\hspace{1em}\=  HAL = 2DArray.new() \\
\>              for d in documents \{ \\
\>\hspace{1em}\=    for i in 1 .. d.len \{ \\
\>\>\hspace{1em}\=      for j in max(1,i-n) .. i-1 \{ \\
\>\>\>\hspace{1em}\=        HAL[d.word(i),d.word(j)] += n+1-(i-j) \\
\>  \}\}\} \\
\>  return HAL \\
end
\end{tabbing}}\end{minipage}
\end{center}
\caption{
  Algorithm to compute the HAL matrix for a collection of
  documents. It is assumed that the documents have been pruned of stop
  words and punctuation.
}
\end{figure}

If word precedence information is considered unimportant the matrix
$S=H+H^T$ denotes a symmetric matrix in which $S[i,j]$ reflects the
strength of association of word $i$ seen in the context of word $j$,
irrespective of whether word $i$ appeared before or after word $j$ in
the context window.  The column vector $S_j$ represents the strengths
of association between $j$ and other words seen in the context of the
sliding window: the higher the weight of a word, the more it has
lexically co-occurred with $j$ in the same context(s).  For example,
table~\ref{table:reagan} illustrates the vector representation for
``Reagan'' taken from a matrix $S$ computed from a corpus of 21578
Reuters news feeds taken from the year 1988.
%[FIXME - how many dimensions in S]
%
\begin{table}[t]
\begin{center}
\begin{tabular}{|p{11cm}|} \hline 
president (5259), administration (2859), trade (1451), house (1426),
budget (1023), congress (991), bill (889), tax (795), veto (786),
white (779), japan (767), senate (726), iran (687), billion (666),
dlrs (615), japanese (597), officials (554), arms (547), tariffs
(536) \dots \\ \hline
\end{tabular}
\end{center}
\caption{Example representation of the word ``Reagan''}
\label{table:reagan}
\end{table}
%

%FIXME
%Maybe use unnormalized weights here as later we compare 
% this to the collapsed vector which is also unnormalized

HAL is an exemplar of a growing ensemble of computational models
emerging from cognitive science, which are generally referred to as
{\em semantic
spaces}~\cite{Article:96:Lund:HAL,Article:98:Burgess:HAL,Article:00:Lowe:SemanticSpace,Article:01:Lowe:SemanticSpace,Article:97:Landauer:LSA,Article:98:Landauer:LSA,Article:97:Patel:SemanticSpace,Article:98:Schuetze:SemanticSpace,Article:99:Levy:SemanticSpace,Article:02:Sahlgren:SemanticSpace}.
Even though there is ongoing debate about specific details of the
respective models, they all feature a remarkable level of
compatibility with a variety of human information processing tasks
such as word association.  Semantic spaces\index{space, semantic}
provide a geometric, rather than propositional, representation of
knowledge.  They can be considered to be approximations of conceptual
space proposed by G\"{a}rdenfors~\cite{Book:00:Gardenfors:ConSpace}.

Within a conceptual space, knowledge has a dimensional structure.  For
example, the property colour can be represented in terms of three
dimensions: hue, chromaticity, and brightness.
G\"{a}rdenfors argues that a
property is represented as a convex region in a geometric space.  In
terms of the example, the property ``red'' is a convex region within
the tri-dimensional space made up of hue, chromaticity and brightness.
The property ``blue'' would occupy a different region of this space.
A domain is a set of integral dimensions in the sense that a value in
one dimension(s) determines or affects the value in another
dimension(s).  For example, the three dimensions defining the colour
space are integral since the brightness of a colour will affect both
its saturation (chromaticity) and hue.  G\"{a}rdenfors extends the
notion of properties into concepts, which are based on domains.  The
concept ``apple'' may have domains taste, shape, colour, etc.  Context
is modelled as a weighting function on the domains, for example, when
eating an apple, the taste domain will be prominent, but when playing
with it, the shape domain will be heavily weighted (i.e., it's
roundness).  One of the goals of this article is to provide both a
formal and operational account of this weighting function.  

Observe
the distinction between representations at the symbolic and conceptual
levels. At the symbolic level ``apple'' can be represented as the
atomic proposition $apple(x)$, however, within a conceptual space
(conceptual level), it has a representation involving multiple
inter-related dimensions and domains.  Colloquially speaking, the
token ``apple'' (symbolic level) is the tip of an iceberg with a rich
underlying representation at the conceptual level.  G\"{a}rdenfors
points out that the symbolic and conceptual representations of
information are not in conflict with each other, but are to be seen as
``different perspectives on how information is described''.

Barwise and Seligman~\cite{Book:97:Barwise:InfoFlow} also propose a
geometric foundation to their account of inferential information
content via the use of real-valued state spaces\index{space, state
space}.  In a state space, the colour ``red'' would be represented as
a point in a tri-dimensional real-valued space.  For example,
brightness can be modelled as a real-value between white (0) and black
(1).  Integral dimensions are modelled by so called observation
functions defining how the value(s) in dimension(s) determine the
value in another dimension.  Observe that this is a similar proposal,
albeit more primitive, to that of G\"{a}rdenfors as the
representations correspond to points rather than regions in the space.

Semantic space models are an approximation of Barwise and Seligman
state spaces whereby the dimensions of the space correspond to
words. A word $j$ is a point in the space.  This point represents the
``state'' in the context of the associated text collection from which
the semantic space was computed.  If the collection changes, the state
of the word may also change.  Semantic space models, however, do not
make provision for integral dimensions.  An important intuition for
the following is the state of a word in semantic space is tied very
much with its ``meaning", and this meaning is context-sensitive.
Further, context-sensitivity will be realized by state changes of a
word.

In short, HAL, and more generally semantic spaces, are a promising,
pragmatic means for knowledge representation based on text.  They are
computational approximations, albeit rather primitive, of
G\"{a}rdenfors' conceptual space.  Moreover, due to their cognitive
track record, semantic spaces would seem to be a fitting foundation
for considering realizing computational variants of human reasoning.
Finally, a semantic space is a real-valued Hilbert space which opens
the door to connections with quantum mechanics.

\section{
  Context effects in Semantic Space
}

Human beings are adept at producing context-sensitive inferences.
Shifts in context effect the inferences made, even to a dramatic
degree. The well known ``Tweety'' problem exemplifies this. A rough
account of this example in terms of G\"{a}rdenfors' model of cognition
is as follows: When given ``Tweety is a bird'', a prototypical concept
of bird is activated within conceptual space and default inferences at
the symbolic level such as ``Tweety flies'' arise as a strong
association is primed between ``Tweety'' and ''flies'' at the
conceptual level. The prototypical ''Tweety'' would be a point in the
centre of a convex region in conceptual space representing birds
(see~\cite{Article:01:Gardenfors:ConSpace},\cite{Book:00:Gardenfors:ConSpace},
p139).  Learning ``Tweety is a penguin'', shifts the representation of
``Tweety'' towards the edge of the region representing birds as
penguins differ significantly to the prototypical bird. As a
consequence the association with ``flies'' diminishes radically and
new associations arise, e.g., with ''Antarctica''. Even though
G\"{a}rdenfors characterizes this type of NMR as being driven by a
change from a general category to a subordinate, we feel that the
associations can be more generally considered as a product of a shift
of context --- in this case the context is being \emph{refined} from broader to
the more specific. Initially the object ``Tweety'' is placed in the
context of the concept ''bird''. The context is then refined to
``penguin'' leading to a change in the associations being primed, and
consequently a change in the inferences being drawn.

The ``Reagan'' example exhibits similar characteristics. The vector
representation given in table~\ref{table:reagan} is almost the
prototypical representation of ``Reagan'' in the context of the underlying
corpus. This is because HAL accumulates the association weights as it
goes along. In fact, the weights in table~\ref{table:reagan} need only
be divided by the frequency of the term ``Reagan'' in the underlying
corpus to produce the vector representing prototypical
``Reagan''. Highly weighted associations in the representation have
the character of being default like - ``Reagan was a president'',
``Reagan had an administration'' etc. Such default associations
reflect the run of the mill presidential Reagan dealing with trade,
budgets, congress etc.

The above two examples exhibit very common, or ``garden'' variety of
practical inference.  In this section, we attempt to provide a formal
account in terms of quantum mechanics (QM).

\subsection{Bridging semantic space and QM}

A semantic space is a vector space\index{space, vector} and these can
be expressed in the notation of quantum mechanics (The following draws
heavily from~\cite{Book:04:Rijsbergen:QM}).

A semantic space $S$ is a $m \times n$ matrix where the columns $\{1,
\ldots, n\}$ correspond to a vocabulary $V$ of $n$ words. A typical
method for deriving the vocabulary is to tokenize the associated
corpus and remove non information bearing words such as ``the'',
``a'', etc.  The letters $u, v, w$ will be used to identify individual
words.

The interpretation of the rows $\{1 \ldots m\}$ depends of the type of
semantic space in question.  For example, table~\ref{table:reaganeg}
illustrates that HAL produces a square matrix in which the rows are
also interpreted as representations of terms.  In contrast, a row in
the semantic space models produced by\index{latent semantic analysis}
Latent Semantic Analysis~\cite{Article:98:Landauer:LSA} corresponds to
a text item, for example, a whole document, a paragraph, or even a
fixed window of text, as above. The value $S[t, w]=x$ denotes the
salience $x$ of word $w$ in text $t$. Information-theoretic approaches
are sometimes use to compute salience. Alternatively, the frequency of
word $w$ in context $t$ can be used.

For reasons of a more straightforward embedding of semantic space into
QM, we will focus on square, symmetric semantic spaces ($m=n$).  A
word $w$ is represented as a column vector in $S$:
\begin{equation} \quad \quad \quad \quad
\ket{w} = \colvec{w}
\end{equation}
The notation on the LHS is called a \emph{ket}, and originates
from quantum physicist Paul Dirac.  Conversely, a row vector $v =
(v_1, \ldots, v_n)$ is denoted by the \emph{bra} \bra{v}.

\index{ket}
\index{bra}

Multiplying a ket by a scalar $\alpha$ is as would be expected:
\begin{equation} \quad \quad \quad
 \alpha\ket{w} =
 \left ( \begin{array}{c}
             \alpha w_1 \\
              \vdots \\
              \alpha w_n
             \end{array}
    \right ) 
\end{equation}
Addition of vectors $\ket{u} + \ket{v}$ is also as one would expect.
In Dirac notation, the scalar product of two $n$-dimensional
real\footnote{QM is founded on complex vector spaces. We restrict our
attention to finite vector spaces of real numbers.} valued vectors $u$
and $v$ produces a real number:
\begin{equation} \quad \quad \quad
\scprod{u}{v} = \sum_{i=1}^n u_i v_i
\end{equation}

The product \ket{u}\bra{u} produces a symmetric matrix.  Vectors $u$
and $v$ are \emph{orthogonal} iff $\scprod{u}{v} = 0$.  Scalar product
allows the length of a vector to be defined: $\lenvec{u} =
\sqrt{\scprod{u}{u}}$.  A vector \ket{u} can be normalized to unit
length ($\lenvec{u}=1$) by dividing each of its components by the
vector's length: $\frac{1}{\lenvec{u}}\ket{u}$.

\index{space, Hilbert}
A Hilbert space is a complete\footnote{The notion of a ``complete''
vector space should not be confused with ``completeness'' in
logic. The definition of a completeness in a vector space is rather
technical, the details of which are not relevant to this account.}
inner product space.  In the formalization to be presented in ensuing
sections, a semantic space $S$ is an $n$-dimensional real-valued
Hilbert space using Euclidean scalar product as the inner product.

A Hilbert spaces allows the state of a quantum system to
be represented.  
It is important to note that a Hilbert
space is an \emph{abstract} state space meaning QM does not prescribe
the the state space of specific systems such as electrons. This is the
responsibility of a physical theory such as quantum electrodynamics.
Accordingly, it is the responsibility of semantic space theory to
offer the specifics: In a nutshell, a ket \ket{w} describes the state
of a word $w$. It is akin to a particle in QM. 
The state of a word changes due to context effects in
a process somewhat akin to quantum collapse\index{quantum collapse}. This in turn bears on
practical inferences drawn due to context effects of word seen
together with other words as described above.

In QM, the state can represent a superposition of potentialities.  By
way of illustration consider the state $\sigma$ of a quantum bit, or
\emph{qubit} as:
\begin{equation} \quad \quad \quad
\ket{\sigma} = \alpha\ket{0} + \beta\ket{1}
\end{equation}
where $\alpha^2 + \beta^2 = 1$.  The vectors \ket{0} and \ket{1}
represent the potentialities, or \emph{eigenstates} of ``off'' and
``on''. Eigenstates \index{eigenstate} are sometimes referred to as \emph{pure}
states. They can be pictured as defining orthogonal axes in a 2-D
plane:
\begin{equation} \quad \quad \quad
 \alpha\ket{0} =
 \left ( \begin{array}{c}
              0 \\
               1
             \end{array}
    \right ) 
\end{equation}
and
\begin{equation} \quad \quad \quad
 \alpha\ket{1} =
 \left ( \begin{array}{c}
              1 \\
               0
             \end{array}
    \right ) 
\end{equation}
The state $\sigma$ is a linear combination of eigenstates. Hard though
it is to conceptualize, the linear combination allows the state of the
qubit to be a mixture of the potentialities of being ``off'' and
``on'' at the same time.  

In summary, a quantum state\index{quantum state} encodes the
probabilities of its measurable properties, or eigenstates. The
probability of observing the qubit being off (i.e., \ket{0} is
$\alpha^2$). Similarly, $\beta^2$ is the probability of observing it
being ``on''.

The above detour into QM raises questions in relation to semantic
space. What does it mean that a word is a superposition - a ``mixture
of potentialities''? What are the eigenstates of a word?

\subsection{Mixed and eigenstates of a word}.

Consider the following traces of text from the Reuters-21578
collection: \emph{
\begin{itemize}
\item President Reagan was ignorant about much of the Iran arms scandal
\item Reagan says U.S to offer missile treaty
\item Reagan seeks more aid for Central America
\item Kemp urges Reagan to oppose stock tax.
\end{itemize}
}

Each of these is a window which HAL will process accumulating weighted
word associations in relation to the word ``Reagan'', say. In other
words, included in the HAL vector for ``Reagan'' are associations
dealing with the Iran-contra scandal, missile treaty negotiations with
the Soviets, stock tax etc.  The point is the HAL vector for
``Reagan'' represents a mixture of potentialities.

Let us now generalize the situation somewhat.  Consider once again the
HAL matrix $H$ computed from the text ``President Reagan ignorant of
the arms scandal''. As mentioned before, $S=H+H^T$ is a symmetric
matrix.  Technically, $S$ is a \emph{Hermitian linear
operator}\index{operator, Hermitian}.  Consider a set of text windows
of length $l$ which are centred around a word $w$. 
Associated with each such text window $j, 1 \leq j \leq m$,
is a semantic space $S_j$.  It is assumed that the semantic space is
$n$-dimensional, whereby the $n$ dimensions correspond to a fixed
vocabulary $V$ as above.  The semantic space around word $w$, denoted
by $S_w$, can be calculated by the sum:
\begin{equation} \quad \quad \quad
  S_w =\sum_{j=1}^kS_j
  \label{eqn:semspace}
\end{equation}

In other words, the semantic space around the word ``Reagan'' is a
summation of $n$-dimensional Hermitian linear operators
\index{operator, Hermitian} computed from text windows centred around
``Reagan''.

\remove{ 
  The mixture can be refined by associating a weight to the
  contributing spaces resulting in the following convex combination.
  \begin{equation} \quad \quad \quad
    S_w  = a_1S_1 + \ldots + a_mS_m
   \label{eqn:sw}
  \end{equation}
  where $a_1 + \ldots + a_m = 1$.
  [FIxME : Minerva 2]
}

In turn, the semantic space of the associated corpus, termed the
\emph{global semantic space}, denoted $\mathcal{S}$ can be considered
as a mixture of the semantic space of the words in the associated
vocabulary $V$:
\begin{equation} \quad \quad \quad
  \mathcal{S} =\sum_{w \in V}S_w
  \label{eqn:globalspace}
\end{equation}

An important intuition drawn from QM is that a word meaning equates with a state.
The state may be \emph{mixed}, that is the state embodies different
potentialities corresponding to different ``senses'' of the word
Reagan.  Here we use the word ``sense'' with some poetic licence, but
we do so deliberately because the ``Reagan'' example is similar to the
case of an ambiguous word.  Consider the word ``suit'' in
isolation. Is it an item of clothing or a legal procedure?  We put
forward the intuition that, both ``Reagan'' or ``suit'' are states
involving mixtures of senses, which parallels the superposition of
eigenstates in the qubit given above.  More formally, let \ket{r} be
the vector representing the state of ``Reagan'' in a semantic space
$S$, and $\{e_1, \ldots, e_k\}$ represent the eigenstates of
$S$, then
\begin{equation} \quad \quad \quad
\ket{r} = \alpha_1\ket{e_1} + \ldots + \alpha_n\ket{e_k}
\end{equation}
where $\alpha_1^2 + \ldots + \alpha_n^2=1$. 
The preceding intuition
connecting word ``meanings" in semantic space to QM seems to have
independently arisen. (See~\cite{Book:04:Widdows:Geometry,Article:03:Widdows:QM,Article:05:Aerts:QMII,Article:04:Aerts:QM}).
The eigenstates define the different senses of the word in
question. In QM terms, these correspond to the eigenstates of ``Reagan''.

In QM,
the interpretation of the eigenstates are clearly grounded, e.g., the
``on'', ``off'' states of a qubit, or the momentum eigenstate of a
particle.  The eigenstates of a word are more subtle.  This subtlety
is not due to subjective interpretations of word meanings.  By using a
semantic space constructed from a corpus of text, the ``meanings''
ultimately are derived from this corpus.  It could be argued that 
such meanings are inter-subjective due to the track record of
semantic space model in replicating human word association norms.  The
subtlety derives more from the range of potential eigenstates.  We
shall see as we go along, however, the state of a word is nevertheless
amenable to a formal treatment.  

\subsection*{Computing eigenstates of a word by Singular Value Decomposition}

\index{singular value decomposition} Singular value decomposition, a
theorem from linear algebra, allows a matrix to be decomposed.  In the
following, many of the technical details of SVD will be skipped, and
only the essential elements will be
presented. See~\cite{Book:93:Golub:Matrix} for a comprehensive
account.
%[FIXME: how is SVD called in QM?]  
As $S_w$ is a symmetric matrix, SVD decomposes it as follows:
\begin{equation} \quad \quad \quad
  S_w = UDU^T
\end{equation}
where $U$ is a $n \times n$ unitary matrix, the columns of which are
the orthonomal basis of $S_w$.  This means, the columns of $U$ are
pairwise orthogonal.  To remain consistent with our notation, the
$i$-th column vector of $U$ will be denoted by $\ket{e_i}$.  Matrix
$D$ is a positive $n \times n$ diagonal matrix, the values of which
are the eigen-values of $S_w$. The value $D[i,i]$ will be denoted
$d_i$.

The spectral decomposition\index{decomposition, spectral} of SVD
allows $S_w$ to be reconstructed, where $k \leq n$:
\begin{eqnarray}
  S_w &= &\sum_{i=1}^k\ket{e_i}d_i\bra{e_i} \\
           & = & \sum_{i=1}^k d_i \ket{e_i}\bra{e_i} \\
           & = & d_1\ket{e_1}\bra{e_1} + \ldots + d_k \ket{e_k}\bra{e_k}
\end{eqnarray}
This shows once again how a word $w$ is a mixture of eigenstates
\ket{e_i}. The eigenvalues are related to the probabilities of the
eigenstates occurring after a quantum measurement\index{measurement,
quantum}.  In the semantic space interpretation, the eigenstates
\ket{e_i} of $S_w$ correspond to the senses of word $w$.

The spectral decomposition of $S_w$ parallels the decomposition of a
density state\index{density state} $\rho$~\cite{Book:04:Rijsbergen:QM}:
\begin{equation} \quad \quad \quad
  \rho = a_1P_1 + \ldots + a_kP_k
  \label{eqn:density}
\end{equation}
where $P_i$ is a projection operator\index{operator, projection} and
$a_1 + \ldots + a_k = 1$. Projection operators, in real valued state
spaces, are idempotent, symmetric matrices.  (Note that $P_i =
\ket{e_i}\bra{e_i}$ is a projection operator)

A \emph{density state}, or \emph{density operator}, or \emph{density
matrix}\index{density matrix} expresses the distribution of quantum
states in an ensemble of particles.  The intuition is that we will run
with is that words are particles, and that context acts like a
``measurement'', e.g., on a particle.  A density matrix is a Hermitian
operator with all its eigenvalues between 0 and 1.

Aerts and Czachor~\cite{Article:04:Aerts:QM} have shown how to render
a semantic space computed by Latent Semantic Analysis into a density
matrix. 
For the purposes of this article, however,
equation~\ref{eqn:density} allows the notion of density matrix to be
directly equated with semantic space when the weights in the space are normalized. This ensures the eigenvalues lie between 0 and 1.

\subsection{An analysis of the eigenstates of ``Reagan''}

Table~\ref{tab: reagan eigenvec} contains the first four eigenstates, or eigenvectors, 
of $S_{\mbox{\sf reagan}}$. The first eigenstate contains all
positive values and can be seen as a kind of average of the space. 
The
subsequent eigenstate has a single positive component and a
collection of negative components. Eigenstates having both positive and
negative components represent two contrasting aspects of the
space. Individual word vectors may project onto either the positive or
the negative portion of each eigenstate. If they were to project
onto both then their projection into the subspace would be small, and
since SVD maximises the average length of the projected vectors, the
negative and positive parts of the eigenstates tend to be in
opposition. The third eigenstate, for example, seems to indicate that
reports about Reagan concerning exports, tariffs and Japan are in
opposition to reports about the senate, vetos and the budget.

It is important to recognize that the eigenstates do not neatly
partition the meanings of Reagan into distinct clusters but rather span
a subspace describing the topics in which Reagan is involved. The
space can be though of as being lumpy but continuous rather than being
due to a small number of discrete and largely disjoint topics.  

In short, the eigenstates computed by SVD do not seem to correspond well
with the intuitively expected eigenstates of the word ``Reagan".

\begin{table}[t]
\begin{center}
\begin{tabular}{|p{11cm}|} \hline 
1: reagan ($0.62$), president ($0.48$), administration ($0.22$), house ($0.17$), trade ($0.15$), congress ($0.11$), budget ($0.11$), bill ($0.10$), veto ($0.10$), white ($0.09$), tax ($0.09$), japan ($0.08$), senate ($0.08$), billion ($0.08$), iran ($0.07$), \\ \hline
2: reagan ($0.74$), \dots

  bill ($-0.04$), congress ($-0.05$), trade
($-0.07$), house ($-0.08$), administration ($-0.23$), president ($-0.55$) \\
\hline
3: japan ($0.25$), trade ($0.25$), japanese ($0.24$), tariffs ($0.21$),
administration ($0.13$), united ($0.11$), sanctions ($0.11$), exports
($0.11$) \dots

tax ($-0.11$), senate ($-0.13$), veto ($-0.14$), budget
($-0.19$), white ($-0.31$), house ($-0.38$) \\ \hline
4: billion ($0.44$), dlrs ($0.37$), dlr ($0.21$), budget ($0.18$), veto
($0.18$), deficit ($0.17$), bill ($0.14$), highway ($0.13$), mln ($0.10$), \dots

conference ($-0.07$), house ($-0.08$), baker ($-0.09$), scandal ($-0.12$),
white ($-0.14$), arms ($-0.24$), iran ($-0.25$) \\ \hline
\end{tabular}\end{center}
\caption{ First four eigenstates of $S_{\mbox{\sf
  reagan}}$. Components are listed in order. Only the largest
  components (by magnitude) are for each eigenstate are shown.
  \label{tab: reagan eigenvec}
}
\end{table}

\subsection{Summary}

As this section has proceeded a sort of duality has emerged.
Initially, the state of a word $w$ was presented as a ket \ket{w} in a
$n$-dimensional semantic (Hilbert) space $S$. 
The ket $\ket{w}$ may represent a
mixture of the senses of the word $w$. Later, a connection was made
between the semantic space $S_w$ constructed around a word $w$ and a
density matrix, a notion from QM.  From a technical point of view,
this is not a problem.  A density matrix can represent both
eigenstates and mixed states: If $\ket{w}$ represents an eigenstate,
then $\ket{w}\bra{w}$ represents the corresponding density matrix.  As
a consequence, the state of a word, whether pure or mixed, can be
represented as density matrix.

However, this technical resolution, does not seem to fully resolve the
perceived duality. For example Widdows~\cite{Article:03:Widdows:QM}
has proposed a quantum of word meanings drawn from semantic space. The
meanings are represented as kets with no recourse to density matrices.
Aerts and Gabora~\cite{Article:05:Aerts:QMII}, on the other hand
employ kets for the pure states of a concept, and a density matrix for
a mixed state of a concept.  It would seem that more research is
needed to resolve this duality.

\section*{Quantum Collapse and Context Effects in Semantic Space}

\index{context effects}
A quantum system is usually not in an eigenstate of whatever
observable (e.g., momentum) is intended to be measured.  However, if
the observable is measured, the state of the system will immediately
become an eigenstate of that observable. This process is known as
\emph{quantum collapse}.

A parallel can be drawn with respect to words in semantic space. When
a word is seen in context, the superposition (mixed) state of the word
collapses onto one of its senses.  The senses of a word are the
observables.  For example, when ``Reagan'' is seen in the context of
``Iran'', the mixture of potentialities of ``Reagan'' collapses onto
the eigenstate representing the sense dealing with the Iran-Contra
scandal.  After collapse, weights of associations to words such as
``Contra'', ``illegal'', ``arms'', ``scandal'', ``sale'' will be high,
whereas before collapse the weights of such associations may have been
weak.  
The highly weighted associations may, for example, ``bubble up" and give rise to defaults at the symbolic level of cognition.
This intuition gives rise to the tantalizing possibility that
context effects within the conceptual level of cognition may be
formalized by quantum collapse.  This change in weighting can be
dramatic and thus produce non-monotonic effects in relation to the weights of associations.
For the moment, the observables can be conceived of as the different
senses of a word. Seeing a word in the context of other word(s) acts
like a ``measurement''. This measurement collapses the word meaning into one of
its potential senses.

The description above of the interaction between context and collapse
is essentially the same as that of Aerts and
Gabora~\cite{Article:05:Aerts:QMII}. They state: ``A state [of a
concept] that is not an eigenstate of the context is called a
potentiality state with respect to this context. The effect of a
context is to change a potentiality state of this context, and this
change will be referred to as collapse''.

\remove{
To each observable parameter of a physical system there corresponds a
Hermitian matrix whose eigenvalues are the possible values that can
result from a measurement of that parameter, and whose eigenvectors
are the corresponding states of the system following a measurement.
}

The context effects that are being considered here are similar to
Aerts and Gabora. For simplicity, the case of a word $v$ seen in the
context of word $u$ will be considered, the prototype of the running
example: ``Reagan'' in the context of ``Iran''.

\subsection{Formalizing context effects by quantum collapse}

Aerts and Gabora~\cite{Article:05:Aerts:QMII} state a measurement in
QM is described by a Hermitian operator $M$.  For the context word
$u$, there is an associated operator $M_u$.  It is assumed that the
state of word $v$ is represented by the \ket{v} drawn from some
density matrix $\rho$.  The parallel with QM is the following --- a
particle (word) $v$ is drawn from a quantum system, the state of which
is $\rho$.  It is subjected to a ``measurement'', which is a product
of word $v$ being seen in the context of word $u$. The state of $v$
collapses as a result.  This intuition is formalized as follows, where
$\ket{v_u}$ denotes the state of word $v$ after collapse:
\begin{equation} \quad \quad \quad
\ket{v_u} = \frac{M_u\ket{v}}{\sqrt{\langle v|M_u|v \rangle}}
\label{eqn:collapse1}
\end{equation}
The value $\sqrt{\langle v|M_u|v\rangle}$ is a normalizing factor.  
One way of inspecting non-montonic effects in relation to associations is 
is simply to compare $\ket{v_u}$ with $\ket{v}$.  
Recall the
ket representation of a word is a vector whose components correspond
to words.  The value $x$ of the component $i$ represents the strength
of association of $\ket{v_u}$ with the $i$-th word of vocabulary $V$.
Examples will follow shortly.

The above equation is a more liberal interpretation of that proposed
by Aerts and Gabora's equation 11 in \cite{Article:05:Aerts:QMII}.
Their equation requires $\ket{v}$ to be a pure state.  Our more
liberal proposal arises from the following intuition: Collapse due to
context may not necessarily result in a pure state. For example,
Reagan's involvement with Iran included the U.S. embassy hostage crisis as
well as the Iran-contra scandal.  Intuitively this phenomena
corresponds to a \emph{partial} collapse of ``Reagan", whereby the resultant state is less
mixed than originally. In other words, the context ``Iran" has not fully
led to a collapse of the ``meaning" of ``Reagan" onto an unambiguous sense.
This phenomenon shows the embedding of semantic space into QM is not
always straightforward.

\subsection{Example: ``Reagan'' in the context of ``Iran''}

To illustrate the effect of equation~\ref{eqn:collapse1}, $\ket{v}$ is
primed to be the state of the word ``Reagan'' extracted from the
density matrix $\rho_{\mbox{\sf Reagan}}$ computed from the Reuters collection.  This
ket represents the prototypical presidential Reagan, and is
illustrated in table~\ref{table:reagan}.  The measurement
operator\index{operator, measurement} $M_u$ is primed as the Hermitian
operator $S_w$ with $w$ equal to the term ``Iran". One interpretation
of the resulting quantum collapse is that it promotes words occurring
in the vicinity of ``Iran'' based on how similar their meaning in the
context of ``Iran'' is to the meaning of the prototypical Reagan.

\remove{
\begin{table}[h]
\begin{center}
\begin{tabular}{|l|l|} \hline 
\textit{Components} & \textit{Value} \\ \hline
iran & 59  \\ 
reagan & 27  \\ 
arms & 21  \\ 
iraq & 12  \\ 
gulf & 12  \\ 
scandal & 10  \\ 
war & 7  \\ 
oil & 7  \\ 
iranian & 7  \\ 
sales & 7  \\ 
... & ... \\ \hline
\end{tabular}
\end{center}
\caption{
  \label{tab:reagancollapse}
  The ``Reagan'' in the context of ``Iran'' 
}
\end{table}

}

\begin{table}[h]
\begin{tabular}{|p{11cm}|} \hline
iran (59), reagan (27), arms (21), iraq (12), gulf (12), scandal (10),
war (7), oil (7), iranian (7), sales (7), house (6), president (5),
attack (5), contra (5), united (5), states (4), white (4), missiles
(4), profits (4), action (3), military (3), officials (3), senate (3),
new (3), tehran (3), shipping (3), news (3), offensive (3), sale (3),
rebels (2), speech (2), secret (2), warned (2), iraqi (2), policy (2),
fighting (2), commission (2), response (2), hussein (2), diversion
(2), major (2), official (2), tower (2), ship (2), denied (2), foreign
(2), deal (2), affair (2), administration (2), saddam (2), \dots \\
\hline
\end{tabular}
\caption{  
  \label{tab:reagancollapse}
  ``Reagan'' in the context of ``Iran".
}
\end{table}

Compare the above weighted associations with those of
table~\ref{table:reagan}.  Observe how the above no longer represent the
prototypical ``Reagan'', but where associations relevant to the
Iran-contra scandal are apparent, e.g., ``scandal'', ``arms'',
``sales'', ``contra''.  Therefore, the Iran-contra sense of Reagan is coming
through.  Also there are prominent associations to ``Iraq'' and
``oil''. These may be related to the sense of ``Reagan'' reflecting
President Reagan's dealings with Iraq during the Iran-Iraq war.
Therefore, table \ref{tab:reagancollapse} seems to reflect two senses
of ``Reagan''.  In other words, the resultant state after collapse is
mixed.  The reason for this is that the context word ``Iran" is also a
mixture of senses.  

Perhaps, for this reason, Aerts and
Gabora~\cite{Article:05:Aerts:QMII} do not directly employ measurement
operator $M$ as a whole, but its spectral decomposition:
\begin{eqnarray*}
  M &=&d_1\ket{e_1}\bra{e_1} + \ldots + d_k \ket{e_k}\bra{e_k}
\end{eqnarray*}
where the projector $P_j= \ket{e_j}\bra{e_j}$.  They refer to
projector $P_j$ as a ``piece of context''.  Take for example,
$u=\{\mbox{Iran}\}$.  This context is a mixture of senses involving
oil trade, Iran-Iraq war, the US embassy siege etc.  The intuition of
the projector $P_j$ is that it represents one of these senses, and
this in turn is a ``piece of context'' which can be substituted in
equation~\ref{eqn:collapse1} instead of $M_u$.

Recourse to ``pieces of context" does not satisfactorily remove an incongruence.
Why is it that when presented with ``Reagan" in the context of ``Iran", most will readily assume the Iran-contra sense, which we argued earlier, is an eigenstate of ``Reagan".
This stands in contrast to the above mixed state of ``Reagan" after collapse illustrated in table~\ref{tab:reagancollapse}.
The progression is as follows. 
Initially the state of ``Reagan" is mixed as reflected by the following ket 
$\ket{r}$ drawn from the Reagan density matrix $\rho_r$.
Assume that the eigenstate of $e_i$ corresponds to the Iran-contra sense of ``Reagan":
\begin{eqnarray*} 
\ket{r} = \alpha_1\ket{e_1} + \ldots + \alpha_i\ket{e_i} + \ldots + \alpha_k\ket{e_k}
\end{eqnarray*}
After collapse due to context ``Iran", the state of ``Reagan" is still mixed but less mixed than before.
The result computed above suggests two senses, denoted $e_i$ and $e_j$:
\begin{eqnarray*} 
\ket{r} = \beta_i\ket{e_i} + \beta_j\ket{e_j}
\end{eqnarray*}
The eigenvalues $\beta_i$ are related to probabilities.
For the sake of argument, let us assume that $\beta_i > \beta_j$.
We speculate that the reason that the eigenstate $e_i$ is assumed by most, is because it is the more probable sense left after collapse.
Bear in mind, these probabilities are furnished by the geometry of the space and
not by a frequentist approach which dominates statistical language processing.

\remove{
The intuition behind context suggests strongly that their effects
should be asymmetric. For example, word $v$ seen in the context of
word $u$, should not necessarily equate with $u$ in the context of
$v$. Table~\ref{tab:irancollapse} illustrates ``Iran'' in the context
of ``Reagan''. The projection has again significantly changed the
vector, this time moving it towards ``Reagan''. Terms such as
``Japan'' have become relatively highly weighted as due to their
similarity in direction to the ``Iran'' vector combined with their
prominence within the ``Reagan'' subspace.

\begin{table}[h]
\begin{tabular}{|p{11cm}|} \hline
reagan (69), president (51), administration (23), iran (17), house
(16), trade (14), congress (10), white (10), budget (9), bill (9),
veto (8), senate (8), japan (8), arms (8), tax (8), billion (7), dlrs
(6), officials (6), japanese (6), united (6), oil (5), states (5),
economic (5), baker (5), new (5), scandal (5), tariffs (5), decision
(4), foreign (4), sales (4), policy (4), secretary (4), washington
(4), conference (4), government (4), gulf (3), plan (3), news (3),
talks (3), chief (3), sanctions (3), nakasone (3), increase (3),
legislation (3), deficit (3), committee (3), action (3), pct (3),
staff (3), contra (3), \dots \\ \hline
\end{tabular}
\caption{  
  \label{tab:irancollapse}
  ``Iran'' in the context of ``Reagan''
}
\end{table}
}
\subsection{Another way to view collapse of word meanings}

Let us assume that before any words are seen or uttered there is a
global density state represented by $\rho_{\mathcal{S}}$, where
$\mathcal{S}$ signifies the global semantic space.  This is akin to a
quantum system with many particles, each particle corresponding to a
word in the vocabulary $V$, which may number in the hundreds of
thousands.  Consider what happens when a word $v$ is expressed in
isolation of other words.  This is transforming a situation without context into one
where the context is simply given by the word $v$.  We contend that
this changes the density state from from $\rho_{\mathcal{S}}$ to
$\rho_v$, which is a subspace of $\rho_{\mathcal{S}}$.  Generalizing
from this intuition leads to the hypothesis that context can be
represented as a projection of a density matrix $\rho$ onto a subspace
represented by another density matrix:
\[
\rho_X = P_X \rho
\]
$P_X$ is a projection operator constructed from one or more context
words represented by $X$.  A word $v$ collapses from $\ket{v} = \rho
\ket{e_i}$ to $\ket{v_X} = \rho_X \ket{e_i} = P_X \rho \ket{e_i} = P_X
\ket{v}$, where $\ket{e_i}$ selects the column from $\rho$ that
corresponds to word $v$. It is curious to note that the $v$ column of
$\rho$, namely $\ket{v}$, is invariant under the transform $P_v$,
because the components of $\ket{v}$ are all drawn from contexts
containing the word $v$, so restricting the context to those containing
the word $v$, i.e. applying $P_v$, doesn't change $\ket{v}$.

The full technical details of $P_X$ still need to be worked out in
relation to semantic space, however the effect of one context word
$X=\{u\}$ can nevertheless be illustrated as $\rho_w$ can be
constructed directly from the Reuters collection via
equation~\ref{eqn:semspace}.

Table~\ref{tab:reagan vector in rho iran} depicts the state of
``Reagan'' in $\rho_{\mbox{\sf Iran}}$ and table~\ref{tab:iran vector
in rho reagan} depicts the state of ``Iran'' in $\rho_{\mbox{\sf
Reagan}}$.  Both tables represent unnormalized kets with the strength
of association to other words represented as values in brackets.

\begin{table}[h]
\begin{tabular}{|p{11cm}|} \hline
iran (827), president (538), arms (430), scandal (208), administration (118),
sales (101), contra (97), speech (91), senate (83), house (80), profits (77), tower (77),
contras (76), commission (73), deal (69), approved (60), conference (58), policy (58),
aid (57), diversion (53), approval (52), fighting (51), poindexter (50), ronald (50),
new (49), rating (48), decision (48), sale (47), funds (47), mistake (47), rebels (46),
wrong (46), investigating (46), denied (45), knew (45), nation (44), news (42),
secret (42), initiative (39), role (39), response (39), pct (38), defense (38),
recollection (38), money (37), congress (37), televised (37), sell (35), adviser (35),
gave (35), \dots \\ \hline
\end{tabular}
\caption{
  \label{tab:reagan vector in rho iran} 
  ``Reagan'' in the context of ``Iran".
}
\end{table}
This ket shows an collapse of the prototypical ``Reagan" onto a state
where associations relevant to the Iran-contra sense are prominently
weighted.  The ket depicted in table~\ref{tab:iran vector in rho
reagan}, however, reflects ``Iran" in the context of ``Reagan".  One
can clearly discern by comparing both kets that context effects are
not symmetric.

\begin{table}[h]
\begin{tabular}{|p{11cm}|} \hline
arms (1522), iraq (1494), gulf (1432), war (939), oil (864), reagan
(827), scandal (639), missiles (620), iranian (594), president (540),
attack (528), offensive (504), sales (463), new (424), shipping (399),
united (396), military (395), states (379), house (370), iraqi (364),
contra (355), action (327), silkworm (291), news (285), hormuz (280),
launched (270), diplomats (268), warned (258), southern (248), sale
(247), major (244), attacked (243), tehran (239), strait (239),
officials (236), kuwait (233), fighting (232), profits (230), north
(225), senate (216), forces (213), foreign (212), washington (203),
shipments (197), soviet (197), strike (196), attacks (193), american
(191), crude (188), mln (185), \dots \\ \hline
\end{tabular}
\caption{
  \label{tab:iran vector in rho reagan} 
   ``Iran'' in the context of ``Reagan".
}
\end{table}

\section{Summary and Outlook}

This article began with speculation that important aspects of human
practical reasoning are manifest within the conceptual level of
cognition referred to as conceptual space. Within conceptual space,
information is represented in a geometric space, and inference has a
associational, rather than a deductive, linear character.  Our
investigation focused on providing a formal account of the
non-monotonic effects on conceptual associations due to context.  Our
aim is to provide the foundations for operational practical reasoning
systems. To this end, the conceptual space was approximated by a
semantic space model which can be automatically derived from a corpus
of text. Within semantic space, words, or concepts, are represented as
vectors in a high dimensional space.  Semantic space models have
emerged from cognitive science and computational linguistics. They
have an encouraging, and at times impressive, track record of
cognitive compatibility with humans across a number of information
processing tasks.  Due to their cognitive credentials semantic space
models would seem to be a fitting foundation for realizing
computational variants of human practical reasoning.  The particular
focus was formalizing the non-monotonic dynamics of associations
within semantic space due to context effects.  Context is a
notoriously slippery notion to pin down. Yet context effects seem to
trigger many garden variety non-monotonic inferences.

It has recently been pointed out in a letter to the editor of a
journal in physics and mathematics that semantic space models bear
some interesting similarities with the framework of quantum mechanics
(QM)~\cite{Article:04:Aerts:QM}.  We have explored the connection
between the two in the light of human practical reasoning and our
intention has been more to provoke thought than provide concrete answers.
It was shown that there is a very close parallel between semantic
space and the notion of a density operator\index{operator, density}
in QM.  In a nutshell, the non-monotonic dynamics of word associations
due to context are formalized by means of the quantum collapse of the
state of a word in semantic space onto a sense which is determined by
context words.  A product of the collapse is a change of state, or
``meaning" of the word. As a consequence, word associations also
change.  QM is one of the few frameworks in which context is neatly
integrated. Essentially, context is something akin to a quantum
measurement which brings about collapse.  We speculate these changes
in word association are the primordial beginnings of non-monotonic
inferences at the symbolic level of cognition.

The embedding of semantic space into QM is not perfect. A summary of
the major problem areas is given as follows:
\begin{itemize}
\item In QM, eigenstates are orthogonal, whereas the senses of a word
need not be.
\item In QM, collapse results in an eigenstate, whereas the collapse
of word meaning in semantic space may be partial.
\end{itemize}
Neither of these problems would seem to fatally undermine further
research.  Aerts , Broekaert and Gabaora ~\cite{Article:05:Aerts:QMa}
go so far to state ``..generalizations of the mathematical formalisms
of quantum mechanics are transferable to the modeling of the creative,
contextual manner in which concepts are formed, evoked, and often
merged together in cognition".  The theory developed in this article
is complemented by realistic illustrations in an operational setting.
The non-monotonic effects witnessed in the illustrations allow for
cautious optimism.

The title of this account includes the phrase ``quantum logic". Where
is the logic?  The phrase ``quantum logic" is promissory. It reflects
our belief that quantum logic, or something akin to it, can be
employed to provide an account of logics of ``down below" meaning
practical reasoning as it is transacted below the symbolic level of
cognition.  It is important to stress that the view of reasoning
presented in this account does not rest on traditional conception of
logic. Gabbay and Woods (\cite{Book:03:Gabbay:Relevance}, p63)
speculate that a logic of ``down below" could be ``a logic of semantic
processing without rules".  We feel that collapse of word meanings in
semantic space falls very much within the ambit of such speculation
and actually reinforces it.

There are yet many stones that need be laid to provide an adequate
bridge between semantic space and quantum logic.  In this regard,
Widdows~\cite{Article:03:Widdows:QM,Book:04:Widdows:Geometry,Article:04:Widdows:QM}
have provided an important contribution with his quantum logic of word
meanings and initial explorations into the lattice structure of vector
subspaces.  Such lattices provide the meeting point for Gabbay and
Engesser's pioneering investigation into the connection between
non-monotonic logic and quantum logic~\cite{Article:02:Gabbay:QM}.

Finally, there is the bigger picture. QM is emerging out of physics and permeating into other areas, for example, information retrieval~\cite{Book:04:Rijsbergen:QM}, human language~\cite{Article:05:Aerts:QMb} and cognition~\cite{Article:05:Aerts:QMa}.
This offers tantalizing possibilities and bizarre implications. 
(See Malin~\cite{Book:03:Malin:QM} for a wonderfully daring view of the philosophical implications of QM).
In terms of semantic space, intriguing questions arise in relation to QM notions such as entanglement. For example, Aerts and Gabora~\cite{Article:05:Aerts:QMII} contend that the pet fish example mentioned in the introductions arises because the concepts ``pet" and ``fish" are entangled. If so, does entanglement manifest in semantic space and can it be exploited in an operational setting? 
Certainly we agree with Aerts and Czachor~\cite{Article:04:Aerts:QM} that the emedding of semantic space models into QM is mostly unexplored. This article documents a tiny exploratory step.

\subsection*{Acknowledgments}

The work reported in this paper has been funded in part by the
Co-operative Research Centre for Enterprise Distributed Systems
Technology (DSTC) through the Australian Federal Government's CRC
Programme (Department of Education, Science, and Training).

In the year 2000, the first author attended a course at ESSLLI on
abduction given by Dov Gabbay and John Woods.  Since then he has been
following their bold and inspiring trail into the ``New Logic".  The
first author thanks Professor Shimon Malin and Dominic Widdows for
their thought provoking discussion and deep insight.

\bibliography{new}

\end{document}